\documentclass[12pt]{article}
\usepackage{epsfig}
\usepackage{subfigure}
\usepackage{amsmath}
\usepackage{amsfonts}
\usepackage{amssymb}
\usepackage{url}
\usepackage{hyperref}
\newcommand{\be}{\begin{equation}}
\newcommand{\ee}{\end{equation}}
\newcommand{\bea}{\begin{eqnarray}}
\newcommand{\eea}{\end{eqnarray}}

\begin{document}
\begin{titlepage}
\begin{flushright}
\today
\end{flushright}
\vspace{4\baselineskip}
\begin{center}
{\Large\bf 
Leptogenesis in a perturbative  SO(10) model}
\end{center}
\vspace{1cm}
\begin{center}
{\large
Tatsuru Kikuchi$^{a,b}$
\footnote{\tt E-mail:tatsuru@post.kek.jp}
}
\end{center}
\vspace{0.2cm}
\begin{center}
${}^{a}$ 
{\small \it Theory Division, KEK,
Oho 1-1, Tsukuba, Ibaraki, 305-0801, Japan}\\
${}^{b}$ {\small \it
Department of Physics, Oklahoma State University, Stillwater, OK 74078, USA}\\
\medskip
\vskip 10mm
\end{center}
\vskip 10mm
%
\begin{abstract}
We consider a phenomenologically viable 
${\rm SO}(10)$ grand unification model which allows perturbative 
calculations up to the Planck scale or the string scale. 
We use a set of Higgs superfields 
${\bf 10 + \overline{16} + 16 + 45}$. 
In this framework, the data fitting of the charged fermion mass matrices 
is re-examined.  
This model can indeed reproduce the low-energy experimental data 
relating the charged fermion masses and mixings. 
As for the neutrino sector, we take the neutrino oscillation data 
as input data to construct right-handed Majorana neutrino mass matrix 
and get a prediction for the physics related to the right-handed 
neutrinos, e.g. the leptogenesis and for the proton decay. 
We propose two kinds of phenomenologically viable model, 
quoted as Model 1 and Model 2. We show that one of the models (Model 2) 
is consistent with all experimental constraints.
\end{abstract}
\end{titlepage}
\section{Introduction}
The supersymmetric (SUSY) grand unified theory (GUT) provides 
an attractive implication for the understandings of the low-energy 
physics. In fact, for instance, the anomaly cancellation between 
the several matter multiplets is automatic in the GUT based on 
a simple gauge group, since the matter multiplets are unified into 
a few multiplets, the experimental data supports the fact of unification 
of three gauge couplings at the GUT scale 
$M_{\mathrm{G}} \sim 2 \times 10^{16}$ [GeV] 
assuming the particle contents of the minimal supersymmetric 
standard model (MSSM) \cite{unification}, and also the right-handed 
neutrino appeared naturally in the ${\rm SO}(10)$ GUT provides a natural 
explanation of the smallness of the neutrino masses through 
the seesaw mechanism \cite{seesaw}. 
While the SUSY GUT has several nice feature described above, 
it should be available up to a scale, several order of magnitude 
above the GUT scale in order to open up a window for really a GUT picture. 
In other words, we require to have perturbative description of the SUSY GUT 
up to a cutoff scale, say the Planck scale or the string scale. 
Once we adopt to have such a landscape, we can not include 
a large dimensional representations like 
${\bf \overline{126}+126}$ of ${\rm SO}(10)$ as a Higgs 
since it causes a strongly asymptotic non-free behavior. 
In addition, a free-field heterotic string theory can never give a 
${\bf \overline{126}+126}$ representations \cite{string}. 
Thus we consider a model based on a set of Higgs with small dimensional 
representations and we explore the minimal choice of the Higgs superfields 
which can accommodate the low-energy experimental data.  
As for the neutrino sector, we take input the neutrino oscillation 
data and get a prediction for the physics related to the right-handed 
neutrinos, e.g. the leptogenesis, etc. 

Here it needs some comments for the alternative approaches are in order.
There could be another possibility which keep the coupling constant perturbative
up to some cut-off scale (that could be the string scale or the Planck scale) 
by including the threshold corrections as shown in Refs. \cite{alternative}.
Those approaches are also interesting as the masses of the colored Higgsinos
can be raised enough to avoid too rapid proton decay without fine-tuning.
However, at least, in this paper, we keep the simple unification picture 
within the framework of the MSSM particle content at low energy.

Now we think about ${\rm SO}(10)$ problems again. 
A very concrete problem we can ask and solve is the followings. 
\vspace{1\baselineskip}
Is there an ${\rm SO}(10)$ model that has:

(1) Gauge couplings unification similar to MSSM RG analysis.

(2) Perturbative unification up to the string or the Planck scale.

(3) Fit all known phenomenology.

\vspace{1\baselineskip}
We think this is a very constrained requirement already. 
To demand perturbative unification up to the string or the 
Planck scale will require the model is pretty ``minimal'' for sure. 
There are two lines of minimal ${\rm SO}(10)$ model. One uses higher dimensional 
representation like ${\bf \overline{126}}$ and renormalizable \cite{babu} 
\cite{matsudaetal} \cite{F-O} \cite{mohapatra}, and the other does not 
use representations higher than ${\bf 54}$ 
representation but non-renormalizable operators \cite{non-ren}. 
In order for satisfying the requirement (2) in the above list of questions, 
we investigate the allowed sets of Higgs superfields by considering 
their contributions to the beta function coefficient. 
In general, we have the RG equation for the unified gauge coupling 
$\alpha_{\rm G}$, 
\be
\frac{1}{\alpha_{\rm G}(\mu)}
=\frac{1}{\alpha_{\rm G}(M_{\rm G})}-
\frac{b}{2 \pi}\log\left(\frac{\mu}{M_{\rm G}}\right). 
\ee
When we set the condition that the coupling constant $\alpha_{\rm G}$ 
blows up at a scale $\mu = \Lambda$ we have the following condition, 
\be
\frac{1}{\alpha_{\rm G}(\Lambda)} \ =\ 0, \quad 
{\mbox{\it i.e. }}, \quad
\frac{2 \pi}{\alpha_{\rm G}(M_{\rm G})} \ =\
b \log\left(\frac{\Lambda}{M_{\rm G}}\right). 
\ee
If we want to have a perturbative description 
up to the (reduced) Planck scale 
$\bar{M}_{\rm pl} =2.4 \times 10^{18}$ GeV, 
we get the upper bound on the coefficient $b$ as 
\be
b \ <\ 34. 
\ee
In general, the coefficient $b$ can be written as 
\be
b \ =\ \sum_{\rm chiral ~multiplet} T({\bf R}) 
- 3 \times \underbrace{8}_{\rm vector ~multiplet} .
\ee
{\sf
\begin{table}
\begin{center}
\begin{tabular}{|c|c|}
\hline
IRREP & T({\bf R}) \\
\hline \hline
{\bf 10} & 1 \\
{\bf 16} & 2 \\
{\bf 45 } & 8 \\
{\bf 54 }& 12 \\
{\bf 120 }& 28 \\
{\bf 126 }& 35 \\
{\bf 210 }& 56 \\
\hline
\end{tabular}
\caption{List of the Dynkin index for the 
${\rm SO}(10)$ irreducible representations up to the 
${\bf 210}$ dimensional one}
\end{center}
\end{table}}
\noindent
When we extract the 3 generation matter contributions we have 
\be
b \ =\ \sum_{\rm Higgs~multiplet} T({\bf R}) + 
3 \times 2 -3 \times 8.
\ee
So the maximal value for the sum of $T({\bf R})$ is given by
\be
\sum_{\rm Higgs~multiplet} T({\bf R}) \ <\ 
34 - 6 + 24 \ =\ 52. 
\ee
For example, a set of Higgs multiplets 
$\left\{\bf 10 \oplus 10^\prime 
\oplus 16 \oplus \overline{16}
\oplus 45 \right\}$ is allowable from perturbative argument, 
\be
\sum_{\rm Higgs~multiplet} T({\bf R}) \ =\ 
\underbrace{1}_{\bf 10\, \oplus\, 10^\prime} \times 2 
+ \underbrace{2}_{\bf 16 \,\oplus\, \overline{16}} \times 2 
+ \underbrace{8}_{\bf 45}  \ =\ 14 \ <\ 52.
\ee
In the former minimal ${\rm SO}(10)$ model including Higgs superpotential 
\cite{FIKMO2}, 
the set of Higgs multiplets is given by 
${\bf 10 \oplus  \overline{126} \oplus 126 \oplus 210 }$, 
thus, in this case, we have a very huge size of the beta 
function coefficient, $b \ =\ 109$. 
In this case, the cutoff scale obtained is very close to the GUT scale 
\bea
\Lambda &=& M_{\rm G} 
\exp \left(\frac{2 \pi}{b \times \alpha_{\rm G}(M_{\rm G})} \right)
\ \simeq\ 4.2 \times M_{\rm G}.
\eea
So this model is severe but not excluded since we have nothing 
definite beyond $M_G$ and even $M_G$ itself. 
Its highly predictivity is deserved to study further. 
However, in this letter, we consider that the existence 
of the string or the Planck scale is an reality that one has to accept.  
So it is actually necessary to include that. 
It is much more important to see if one can get perturbative 
unification up to the string or the Planck scale 
even when higher dimensional operators are included.  
While there are {\it a priori} many choices for the Higgs superfields, 
the predictivity and the perturbativity may pick up some representations.  
And we explore the minimal choice of the Higgs superfields 
which can accommodate the low-energy experimental data.  
As for the neutrino sector, we take input the neutrino oscillation 
data and get a prediction for the mass matrix 
of the right-handed neutrinos. 

\section{Fermion mass matrices }
We can consider some possible choices for 
Higgs representations to be introduced in our ${\rm SO}(10)$ model.  
In any case, they should accomplish the following tasks: 
(1) it can break the ${\rm SO}(10)$ gauge symmetry down to the standard model one.  
(2) it can reproduce all the fermion mass matrices being realistic.  
For the task (1), we give some examples, 
${\bf 45 + 54}$, ${\bf 16 + \overline{16} + 45}$ etc.  
Considering the second task and the fact that top Yukawa coupling 
is of order one, it is necessary to introduce, at least, 
one ${\bf 10}$ Higgs representation which can provide 
a renormalizable Yukawa coupling.  
Note that ${\bf 126}$ Higgs representation is excluded by our criteria 
of the perturbative unification, since ${\bf 126 + \overline{126}}$ 
contributes $b_{\mathrm{Higgs}} = 70$.  
Furthermore, in order to incorporate right-handed neutrino 
Majorana masses, 
we need ${\bf \overline{16}}$ Higgs, and consider 
the superpotential 
\bea 
W= \frac{1}{M}\,Y_{\overline{16}}^{ij} {\bf 16}_i {\bf 16}_j 
{\bf \overline{16}}_H {\bf \overline{16}}_H, 
\label{Maj}
\eea
where $M$ is the cutoff scale of our model, 
the Planck scale or the string scale, for example.  
In this paper, we consistently assume that only the singlet 
in ${\bf \overline{16}}_H$ ($\widetilde{\bar{\nu}}_H$) 
and ${\bf 16}_H$ ($\widetilde{\nu}_H$) 
can achieve their VEVs and the doublets never get their VEVs. 
This assumption is essential to write down the GUT mass relations 
for the charged fermions. 
Then we may define our ``minimal model'' as the one 
which satisfies all the above requirements and contributes 
$b_{\mathrm{Higgs}}$ as small as possible.  
The most reasonable candidate is the choice with 
${\bf 10 + 16 + \overline{16} + 45}$ Higgs multiplets.  

With these Higgs multiplets, the superpotential 
relevant to the fermion mass matrices is given by 
(up to dimension 5 terms) 
\bea 
W &=&  Y_{10}^{ij} {\bf 16}_i {\bf 16}_j {\bf 10}_H 
\ +\  \frac{1}{M}\,Y_{45}^{ij} {\bf 16}_i {\bf 16}_j {\bf 10}_H {\bf 45}_H 
\label{line1}  \\
&+&  g_{i} {\bf 16}_i {\bf \overline{16}}\, {\bf 45}_H 
\ + \  h_{i} {\bf 16}_i {\bf 16}\, {\bf 10}_H 
\label{line2}
\\
&+& \tilde{g} \,{\bf 16} \,{\bf \overline{16}}\, {\bf 45}_H 
\ + \  \tilde{h} \,{\bf 16}\, {\bf 16}\, {\bf 10} 
\ + \  \tilde{f} \,{\bf \overline{16}}\, {\bf \overline{16}}\, {\bf 10}_H 
\label{line3}
\\
&+&
M_{10} {\bf 10}_H^2 
\ +\
M_{16} {\bf 16}_H {\bf \overline{16}}_H 
\ +\
M_{45} {\bf 45}_H^2 
\ +\
\lambda {\bf 16}_H {\bf \overline{16}}_H {\bf 45}_H  \;, 
\label{line4}
\eea
where the Yukawa coupling matrices $Y_{10}$ is symmetric, 
while $Y_{45}$ is antisymmetric. 
Here, an extra vector like $\{{\bf 16} + \overline{\bf 16}\}$-multiplet with no subscript represents
an extra matter multiplet, whose matter parity is assigned to be odd, so that the Yukawa coupling
given in Eqs.~(\ref{line1})-(\ref{line4}) becomes invariant under the matter parity.

Note that the ${\bf 16}_i$ and ${\bf 16}$ do mix with each other. 
So, in general, the Yukawa couplings can be written as 
\bea
W &=&
\left(
\begin{array}{ccc}
{\bf 16}_i, &
{\bf 16}, & 
{\bf \overline{16}}
\end{array}
\right)
\left(
\begin{array}{ccc}
Y_{ij} {\bf 10}_H & (h_{i}/2)\, {\bf 10}_H & (g_{i}/2)\, {\bf 45}_H \\
(h_{j}/2)\, {\bf 10}_H & \tilde{h} {\bf 10}_H & (\tilde{g}/2)\, {\bf 45}_H \\
(g_{j}/2)\, {\bf 45}_H & (\tilde{g}/2)\, {\bf 45}_H & \tilde{f} {\bf 10}_H 
\end{array}
\right)
\left(
\begin{array}{c}
{\bf 16}_j \\
{\bf 16} \\
{\bf \overline{16}}
\end{array}
\right) 
\nonumber\\
&=&
\left(
\begin{array}{ccc}
{\bf 16}_i, &
{\bf 16}, & 
{\bf \overline{16}}
\end{array}
\right)
\left(
\begin{array}{ccc}
{\cal O } (M_W) & {\cal O } (M_W) & {\cal O } (M_{\rm G}) \\
{\cal O } (M_W) & {\cal O } (M_W) & {\cal O } (M_{\rm G}) \\
{\cal O } (M_{\rm G}) & {\cal O } (M_{\rm G}) & {\cal O } (M_W) 
\end{array}
\right)
\left(
\begin{array}{c}
{\bf 16}_j \\
{\bf 16} \\
{\bf \overline{16}}
\end{array}
\right), 
\label{55}
\eea
where we provided the VEVs to the Higgs, 
$\left<{\bf 10}_H \right> \sim {\cal O } (M_W)$ and 
$\left<{\bf 45}_H \right> \sim {\cal O } (M_{\rm G})$ 
and this mass matrix has only one weak scale mass ${\cal O } (M_{\rm W})$ 
and the remaining two are of order the GUT scale ${\cal O } (M_{\rm G})$. 
Then three lighter modes in the multiplet ${\bf 16}_\alpha = 
\left({\bf 16}_j, \,{\bf 16},\,{\bf \overline{16}} \right)_\alpha$ 
$(\alpha = 1,2,3,4,5)$ are identified with the ${\bf 16}$-multiplet 
including the usual quarks and leptons. 
The light modes $({\bf 16}_i')~(i=1,2,3)$ in the mass matrix of Eq.~(\ref{55}), 
which contains the usual quarks and leptons can be written as 
the following linear combinations (up to some normalization factor):
\bea
{\bf 16}_1' &=& - g    {\bf 16}_1 + g_1 {\bf 16} \;,
\nonumber\\
{\bf 16}_2' &=& - g_3 {\bf 16}_2 + g_1 {\bf 16}_3 \;,
\nonumber\\
{\bf 16}_3' &=& - g_2 {\bf 16}_3 + g_1 {\bf 16}_2 \;.
\eea
The remaining two eigenstates become heavy of order $M_{\rm G}$,
so the low energy effective theory can indeed become the MSSM
without exotics.
Then the extra vector-like matter ${\overline{q}_L^\prime \oplus q_L^\prime}$, 
${\overline{\ell}_L^\prime \oplus \ell_L^\prime}$, 
{\it etc.} would decouple from the low-energy effectve field theory. 
After giving a VEV to the ${\bf 45}_H$ Higgs multiplet, 
we have the dimension 5 neutrino mass operator Eq. (\ref{Maj}) 
and also the dimension 6 operator contributing to the 
charged fermion mass matrices 
\be
W = \frac{1}{M}\,Y_{\overline{16}}^{ij} {\bf 16}_i {\bf 16}_j 
{\bf \overline{16}}_H {\bf \overline{16}}_H
+ \left(\frac{g_{\left\{i \right.} h_{\left.j\right\}}}
{\tilde{g} M_{16}^2} \right)
{\bf 16}_i {\bf 16}_j {\bf 10}_H {\bf 45}^2_H. 
\label{dim6}
\ee
In the followings, we consider two cases for the description 
of the charged fermion mass matrices. 
One is to consider the situation in which the second term 
in Eq. (\ref{line1}) has larger contribution than Eq. (\ref{dim6}). 
Another one is the opposite case in which the dominant term 
is Eq. (\ref{dim6}) rather than the second term in Eq. (\ref{line1}). 

In the former case, Eq.~(\ref{line1}) corresponds to 
the Dirac mass matrices of quarks and leptons. 
Note that this example is obviously unrealistic 
since it predicts the Kobayashi-Maskawa matrix being unity.  
This is because the Higgs doublets in Eq.~(\ref{line1}) 
is the same, and as result 
the up-type quark mass matrix is proportional to 
the down-type quark mass matrix. 
One way to avoid this problem is to introduce 
new ${\bf 10}$ Higgs and a symmetry which allow the superpotential 
such as 
\bea 
W \ =\  Y_{10}^{ij} {\bf 16}_i {\bf 16}_j {\bf 10}_1          
\ +\  \frac{1}{M}\,Y_{45}^{ij} {\bf 16}_i {\bf 16}_j {\bf 10}_2 {\bf 45}_H , 
\label{example}
\eea
where ${\bf 10}_1$ and ${\bf 10}_2$ are two Higgs multiplets 
of ${\bf 10}$ representation.  
Since ${\bf 10} \times {\bf 45} \supset {\bf 10 + 120 + 320}$, 
this system is effectively the same as the one 
with ${\bf 10 + 120}$ Higgs multiplets.  

In the latter case, 
the system is completely equivalent to the minimal ${\rm SO}(10)$ 
model which uses ${\bf 10 + \overline{126}}$ Higgses 
w.r.t. the charged fermion sector since 
${\bf 10} \times {\bf 45}^2 \supset {\bf 126 + \overline{126}}$. 
In this case we need only one ${\bf 10}$ multiplet 
to reproduce realistic fermion mass matrices. 

\subsection*{Model 1}
In the following notation, we use ${\bf 120}$ Higgs, for simplicity.
This model has been studied in \cite{Chang:2004pb}, in detail,
and we give a very brief review of this model.

The Yukawa couplings relevant to the Dirac mass matrices 
are given by 
\bea 
 W= Y_{10}^{ij} {\bf 16}_i {\bf 16}_j {\bf 10}_H 
+Y_{120}^{ij} {\bf 16}_i {\bf 16}_j {\bf 120}_H ,  
\eea  
where $Y_{10}$ and $Y_{120}$ are symmetric and anti-symmetric, 
respectively.  Number of free parameters in the Yukawa matrices 
is found to be $3+3 \times 2 =9$ in total. 
Both of the Higgs multiplets ${\bf 10}_H$ and ${\bf 120}_H$ include 
a pair of Higgs doublets in the MSSM decomposition.  
At low-energy after the GUT symmetry breaking, the superpotential 
leads to
\footnote{
In general, there could be another contribution to the mass matrices from bi-doublet 
$({\bf 1}, {\bf 2}, {\bf 2}) \subset {\bf 120}$ under $G_{422} = {\rm SU}(4) \times {\rm SU}(2)_L \times {\rm SU}(2)_R$.
In this paper, we assumed that it does not contribute to the fermion mass matrices,
just for simplicity.
}
\bea 
 W&=& (Y_{10}^{ij} H_{10}^u + Y_{120}^{ij} H_{120}^u) u^c_i q_j 
+ (Y_{10}^{ij} H_{10}^d + Y_{120}^{ij} H_{120}^d) d^c_i q_j  \nonumber \\
&+& (Y_{10}^{ij} H_{10}^u -3  Y_{120}^{ij} H_{120}^u) N_i {\ell}_j  
+ (Y_{10}^{ij} H_{10}^d -3 Y_{120}^{ij} H_{120}^d) e^c_i {\ell}_j , 
\eea  
where $H_{10}$ and $H_{120}$ correspond to the Higgs doublets 
in ${\bf 10}$ an ${\bf 120}$ Higgses, which originate 
${\bf 10}_1$ and ${\bf 10}_2$ in the original superpotential 
Eq.~(\ref{example}).  
The factor $3$ in the lepton sector is the results from 
the VEV of ${\bf 45}$ Higgs in the $B-L$ direction.  
Providing VEVs for all the Higgs doublets, 
the Dirac mass matrices are obtained as 
\bea 
W_{mass} = M_u^{ij} u^c_i q_j + M_d^{ij} d^c_i q_j  
         + M_D^{ij}  N_i {\ell}_j + M_e^{ij}  e^c_i {\ell}_j , 
\eea
where 
\bea 
M_u &=& c_{10} M_{10} +c_{120} M_{120},  \nonumber \\ 
M_d &=&  M_{10} +  M_{120},  \nonumber \\ 
M_D &=& c_{10} M_{10} -3 c_{120} M_{120},  \nonumber \\ 
M_e &=&  M_{10} -3  M_{120}  . 
\label{massmatrix}
\eea 
Here $c_{10} = \langle H_{10}^u \rangle/\langle H_{10}^d \rangle$ 
and  $c_{120} = \langle H_{120}^u \rangle/\langle H_{120}^d \rangle$ 
are complex parameters in general.  
Now number of free parameters in terms of mass matrices are found 
to be $3+ 3 \times 2 + 2\times 2 =13$.  
This relation leads to a relation among the mass matrices of 
up- and down-type quarks and charged lepton such as 
\bea 
M_e = c_d \left( M_d + \kappa M_u \right),  
\label{GUTrelation}
\eea
where 
\bea
c_d = - \frac{3 c_{10} +c_{120}}{c_{10} - c_{120}}, \nonumber \\
\kappa = - \frac{4}{3 c_{10} + c_{120}}  . 
\eea
Note that this GUT relation holds at the GUT scale.  
By the same methods in \cite{F-O}, we will solve the relation
and find mass matrices compatible to the low-energy data 
of the fermion mass matrices.  

Without loss of generality, we can begin with the basis 
where $M_u$ is real and diagonal, $M_u = D_u$. 
Here we assume $M_d$ to be the hermitian matrix, 
and the number of parameters are reduced to $3+3+4=10$. 
Therefore it can be diagonalized by an unitary matrix, 
$M_d = V_{\rm KM} \,D_d \,V_{\rm KM}^\dagger$. 
Considering the basis-independent quantities, 
$\mathrm{tr}\left(M_e \right)$, 
$\mathrm{tr}\left(M_e^2 \right)$ 
and $\mathrm{det}\left(M_e \right)$, 
and eliminating $c_d$, we obtain two independent equations,  
\begin{eqnarray}
\left(
\frac{\mathrm{tr} (\widetilde{M_e} )}
{m_e + m_{\mu} + m_{\tau}} \right)^2
&=& 
\frac{\mathrm{tr} (\widetilde{M_e}^2 )}
{m_e^2 + m_{\mu}^2 + m_{\tau}^2},
\\
\left( \frac{\mathrm{tr} (\widetilde{M_e} )}
{m_e + m_{\mu} + m_{\tau}} \right)^3
&=&
\frac{\mathrm{det} (\widetilde{M_e} )}
{m_e \; m_\mu \; m_\tau},
\end{eqnarray}
where $\widetilde{M_e} \equiv 
V_{\rm KM} \, D_d \, V_{\rm KM}^\dagger + \kappa D_u$. 
With input data of six quark masses, 
 three angles and one CP-phase in the CKM matrix 
 and three charged lepton masses, 
 we can solve the above equations 
 and determine $\kappa$ and $c_d$. 
The original basic mass matrices, $M_{10}$ and $M_{120}$, 
 are described by 
\begin{eqnarray}
M_{10} 
&=& 
\frac{3+ c_d}{4} \;
V_{\rm KM} , D_d \, V_{\rm KM}^\dagger
+ \frac{c_d\; \kappa}{4} \;D_u, 
\\ 
M_{120} &=&
\frac{1- c_d}{4} \;
V_{\rm KM} \, D_d \, V_{\rm KM}^\dagger
- \frac{c_d\; \kappa}{4} \;D_u. 
\end{eqnarray} 
Now $M_{10}$ and $M_{120}$ are completely determined 
with the solutions $c_d$ and $\kappa$ 
determined by the GUT relation. 
\footnote{
Although the solution is found in \cite{matsudaetal}, 
the GUT relation are applied at the electroweak scale.  
The analysis incorporating renormalization effects was done as in \cite{F-O}.}
It should be checked that $Y_{120}^{ij}$ should be much 
smaller than one, 
since $Y_{120}^{ij} = Y_{45}^{ij} \langle {\bf 45} \rangle/M $.  
In the next section, 
we will perform a numerical analysis by using the same method as 
\cite{F-O}. 

\subsection*{Model 2}
The Yukawa couplings relevant to the Dirac mass matrices 
are given by 
\bea 
 W= Y_{10}^{ij} {\bf 16}_i {\bf 16}_j {\bf 10}_H 
+Y_{126}^{ij} {\bf 16}_i {\bf 16}_j {\bf \overline{126}}_H ,  
\eea  
where both $Y_{10}$ and $Y_{126}$ are symmetric, and 
the number of free parameters in the Yukawa matrices 
is found to be $3+6 \times 2 =15$ in total. In this model, 
the GUT relation is the same as the one in Eq. (\ref{GUTrelation}). 
But in this case, the coefficients $c_{10}$ and $c_{126}$ 
are complex numbers, and then the analysis for finding 
a solution is somewhat restrictive. 
Notice that the right-handed neutrino Majorana mass matrix 
is completely free in this model as depicted in Eq. (\ref{Maj}), 
thus we can reproduce the neutrino oscillation data. But even in this 
case, since we know the neutrino Dirac mass matrix, we can get 
a prediction for the quantity relating the leptogenesis. 

Now as in the same way in Model 1, 
considering the basis-independent quantities,  
$\mathrm{tr} [M_e^\dagger M_e ]$, 
$\mathrm{tr} [(M_e^\dagger M_e)^2 ]$ 
and $\mathrm{det} [M_e^\dagger M_e ]$, 
and eliminating $|c_d|$, we obtain two independent equations,  
\begin{eqnarray}
\left(
\frac{\mathrm{tr} [\widetilde{M_e}^\dagger \widetilde{M_e} ]}
{m_e^2 + m_{\mu}^2 + m_{\tau}^2} \right)^2
&=& 
\frac{\mathrm{tr} [( \widetilde{M_e}^\dagger \widetilde{M_e} )^2 ]}
{m_e^4 + m_{\mu}^4 + m_{\tau}^4},
\\ 
\left( \frac{\mathrm{tr} [\widetilde{M_e}^\dagger \widetilde{M_e} ]}
{m_e^2 + m_{\mu}^2 + m_{\tau}^2} \right)^3
&=&
\frac{\mathrm{det} [\widetilde{M_e}^\dagger \widetilde{M_e} ]}
{m_e^2 \; m_\mu^2 \; m_\tau^2},
\end{eqnarray}
where $\widetilde{M_e} \equiv U^* \, D_d \, U^\dagger + \kappa D_u$. 
With input data of six quark masses, 
three angles and one CP-phase in the CKM matrix 
and three charged lepton masses, 
we can solve the above equations 
and determine $\kappa$ and $|c_d|$, 
but five parameters, the phase of $c_d$ \cite{F-O} and 
the phases in $U$, except for the CKM phase are undetermined. 
\footnote{In this paper, we ignore the additional phases in $U$ 
except for the KM phase $\delta$, and thus hereafter 
we denote $U$ just as $V_{\rm KM}$. }
The original basic mass matrices, $M_{10}$ and $M_{126}$, 
are described by 
\begin{eqnarray}
M_{10} 
&=& 
\frac{3+ |c_d| e^{i \sigma}}{4} 
V_{\rm KM}^* \, D_d \, V_{\rm KM}^\dagger
+ \frac{|c_d| e^{i \sigma} \kappa}{4} D_u, 
\label{M10}  \\ 
M_{126} &=& 
\frac{1- |c_d| e^{i \sigma}}{4} 
V_{\rm KM}^* \, D_d \, V_{\rm KM}^\dagger
-\frac{|c_d| e^{i \sigma} \kappa}{4} D_u. 
\label{M126} 
\end{eqnarray}
By giving a solution for $|c_d|$ and $\kappa$, we 
can determine all the Dirac mass matrix as a function of a phase $\sigma$. 
As a result, we can obtain a prediction for the leptogenesis 
parameters. 

\section{Numerical analysis and results} 
Now we are ready to perform all the numerical analysis. 
We first show the solution in Model 1. 
In the following analysis, we assume all the charged fermion 
mass matrices to be hermitian.  
We take input the absolute values of the fermion masses at $M_Z$ 
as follows (in GeV) \cite{PDG}: 
\bea
m_u &=& 0.00233, \quad m_c = 0.677, \quad m_t = 176, \nonumber\\
m_d &=& 0.00469, \quad m_s = 0.0934, \quad m_b = 3.00, \nonumber\\
m_e &=& 0.000487, \quad m_{\mu} = 0.103, \quad m_{\tau} = 1.76.  
\eea
Here the signs of the input fermion masses have taken as 
$\left(m_u, m_c, m_t \right) = \left( -,-,+ \right)$ 
and 
$\left(m_d, m_s, m_b \right) = \left( -,-,+ \right)$.  
And for the CKM mixing angles and a CP-violating phase 
in the "standard"  parameterization, 
we input the center values measured by experiments as follows: 

\be
s_{12} = 0.2229, \quad s_{23} = 0.0412, \quad s_{13} = 0.0036, \quad
\delta = \pi/3 \,\, [{\mathrm{rad}}]. 
\ee
Since it is very difficult to search all the possible parameter region 
systematically, we present only reasonable results we found.  
In the following, we show our analysis in detail 
in the case $\tan \beta=50$.  
After the RG running, we get the values at the GUT scale are use them 
as input parameters in order to solve the GUT relation of 
Eq. (\ref{GUTrelation}).  

Now for simplicity, we assume the matrices $M_{10}$ and $M_{120}$ 
in Eq. (\ref{massmatrix}) to be hermitie.  
Thus the coefficients $c_{d}$ and $\kappa$ are real parameters.  
Then we find a solution 
\begin{eqnarray}
c_d &=& - 11.1923, \nonumber\\ 
\kappa &=& -0.01433. 
\end{eqnarray}
This means we could reproduce the low-energy experimental data 
related to the charged fermion sector. 
As discussed in the previous section 
the Yukawa coupling matrices, $Y_{10}$ and $Y_{120}$, 
are related to the corresponding mass matrices 
$M_{10}$ and $M_{120}$ such that 
\bea
Y_{10} &=& \frac{c_{10}}{\alpha^u  v \sin{\beta}} M_{10}, 
\nonumber\\
Y_{120} &=& \frac{c_{120}}{\beta^u  v \sin{\beta}} M_{120}. 
\eea
Here $\alpha^u$ and $\beta^u$ are the Higgs doublet mixing parameters 
introduced in \cite{Detailed}, which are restricted in the range 
$|\alpha^u|^2 +|\beta^u|^2 \leq 1$. Although these parameters are 
irrelevant to fit the low-energy experimental data of the fermion 
mass matrices, there are theoretical lower bound on them in order 
for the resultant Yukawa coupling constant not to exceed the 
perturbative regime. 
Now we show one example of the Yukawa coupling matrices 
with fixed $\alpha^u = 0.707$, 
\be
Y_{10}
=
\left(
\begin{array}{ccc}
\begin{array}{c}
0.00304  \\
0.00644  \\
-0.00236 \\
\end{array}
\begin{array}{c}
0.00644  \\
0.0235 \\
-0.0502 
\end{array}
\begin{array}{c}
-0.00236  \\
-0.0502  \\
0.761 
\end{array}
\end{array}
\right),
\ee
\be
Y_{120}
=
\left(
\begin{array}{ccc}
\begin{array}{c}
0  \\
0.000170 \, i \\
0.00557 \, i 
\end{array}
\begin{array}{c}
-0.000170 \, i \\
0 \\
0.0000226 \, i 
\end{array}
\begin{array}{c}
- 0.00557 \, i \\
- 0.0000226 \, i \\
0
\end{array}
\end{array}
\right). 
\ee
For the proton decay analysis, we search for the parameters which 
cancel the proton decay rate, completely. 
In the followings, we restrict the region of the parameters 
in the range $(\alpha^u)^2 +(\beta^u)^2 = 1$ 
(we assume $\alpha^u$ and $\beta^u$ real for simplicity). 
Then we actually find a solution that cancels the proton decay 
$p \to K^+ \overline{\nu}$. The resultant $2 \times 2$ 
color-triplet mass matrix \cite{Detailed} has the following form: 
\be
M_C = M_{\rm G} {\bf I}_2 \times 
\left(
\begin{array}{cc}
\begin{array}{c}
-0.0814 + 0.113 \, i \\
0.872 - 0.470 \, i 
\end{array}
\begin{array}{c}
-0.872 - 0.470 \, i \\
-0.0814 - 0.113 \, i 
\end{array}
\end{array}
\right).
\ee
For about Model 2, we have already found a solution 
in \cite{Detailed}. Therefore we just present the result here 
\begin{eqnarray}
& \kappa =
-0.00675 + 0.000309\, i \;,   \nonumber\\
& |c_d| = 5.99  \; , & 
\end{eqnarray}
with $\tan\beta = 2.5$. 

\section{Neutrino oscillation data}
In our scheme adopted here, the right-handed neutrino mass matrix is 
left free from fitting the charged fermion data.  
We know the definite structure for the Dirac neutrino mass matrix.  
Therefore, by making use of the neutrino oscillation data, 
we can give a definite prediction for the right-handed neutrino 
mass matrix.  
\bea
M_R &=& M_D \;M_\nu^{-1} \;M_D^{T}
\nonumber\\
    &=& 
M_D\;
U_{\mathrm{MNS}}^{\dagger} \;{\mathrm{diag}} 
(m_1^{-1},m_2^{-1},m_3^{-1})\; U_{\mathrm{MNS}}^{\ast} \;
M_D^{T} .
\eea
Here $U_{\mathrm{MNS}}$ is the Maki-Nakagawa-Sakata (MNS) lepton 
mixing matrix, and in the standard parametrization, it can be 
written as 
\bea
U_{\mathrm{MNS}}
&=&
\left(
\begin{array}{ccc}
c_{13}c_{12} & c_{13}s_{12} & s_{13}e^{-i\delta} \\
-c_{23}s_{12}-s_{23}c_{12}s_{13} e^{i\delta}
&c_{23}c_{12}-s_{23}s_{12}s_{13} e^{i\delta} 
&s_{23}c_{13} \\
s_{23}s_{12}-c_{23}c_{12}s_{13} e^{i\delta}
 & -s_{23}c_{12}-c_{23}s_{12}s_{13} e^{i\delta} 
& c_{23}c_{13} \\
\end{array}
\right)
\nonumber\\
&\times&
{\mathrm{diag}}(1, e^{i \beta}, e^{i \gamma}), 
\eea
where $s_{ij}:= \sin \theta_{ij}$, $c_{ij}:=\cos \theta_{ij}$, 
$\delta$, $\beta$, $\gamma$ are the Dirac phase and the Majorana 
phases, respectively.  
Now we take input the center values of a global analysis 
for the neutrino oscillation parameters after the recent KamLAND data 
\cite{petcov}.  
\footnote{Our convention is $\Delta m_{ij}^2 = m_i^2 -m_j^2$.  }
\bea
\Delta m^2_{\oplus} &=& 
\Delta m^2_{21} 
\ =\ 2.1 \times 10^{-3}\;\; {\mathrm{eV}}^2,
\nonumber\\
\sin^2 \theta_{\oplus} &=& 0.5,
\nonumber\\
\Delta m^2_{\odot} &=& 
\left| \Delta m^2_{31} \right| 
\ =\ 8.3 \times 10^{-5}\;\; {\mathrm{eV}}^2,
\nonumber\\
\sin^2 \theta_{\odot} &=& 0.28, 
\nonumber\\
|U_{e3}| &<& 0.15.
\eea
Note that we can take both signs of $\Delta m^2_{31}$, 
$\Delta m^2_{31} > 0$ or $\Delta m^2_{31} < 0$.  
The former is called ``normal hierarchy'' (NH), the latter is called 
``inverted hierarchy'' (IH), and we study both cases.  
Writing the lightest neutrino mass eigenvalue as $m_\ell$, 
we can write the mass eigenvalues as 
\bea
m_1 &=& m_\ell, 
\nonumber\\
m_2 &=& \sqrt{m_\ell^2 + \Delta m^2_{\oplus}}, 
\nonumber\\
m_3 &=& \sqrt{m_\ell^2 + \Delta m^2_{\oplus} + \Delta m^2_{\odot}}, 
\eea
for the case of NH, and 
\bea
m_1 &=& \sqrt{m_\ell^2 + \Delta m^2_{\oplus} + \Delta m^2_{\odot}}, 
\nonumber\\
m_2 &=& \sqrt{m_\ell^2 + \Delta m^2_{\oplus}},
\nonumber\\
m_3 &=& m_\ell, 
\eea
for the case of IH.  

Because we know the Dirac mass matrix $M_D$, by inputting 
the above neutrino oscillation data, we can predict the mass matrix 
of the right-handed neutrinos as a function 
of Majorana phases, $\beta$ and $\gamma$.  As a result, we can 
have a prediction on the phenomena relating the right-handed neutrinos, 
such as the leptogenesis, lepton flavor violating processes (LFV), etc.  

Now we turn to the discussions about the baryon asymmetry of 
the universe based on the leptogenesis scenario \cite{FY}.  
In the leptogenesis scenario, lepton asymmetry is generated by 
the out-of-equilibrium decays of the right-handed neutrinos.  
The lepton asymmetry of the right-handed neutrino $N_i$ 
is defined as 
\be
\epsilon_i =
\frac{\Gamma(N_i \rightarrow \ell \;H) 
- \Gamma(N_i \rightarrow \ell^c \; H^\dagger)}
{\Gamma(N_i \rightarrow \ell \; H) 
+ \Gamma(N_i \rightarrow \ell^c \; H^\dagger)}. 
\ee
At tree level, the decay width of $N_i$ is given by 
\be
\Gamma_{i} = \frac{(M_D M_D^{\dag})_{ii}}{8\pi v^2 \sin^2 \beta} M_i. 
\ee
The CP asymmetry is produced for the first time on one-loop level as 
\be
\epsilon_i = 
-\frac{1}{8\pi v^2 \sin^2 \beta}\frac{1}{(M_D M_D^{\dag})_{ii}}
\sum_{j\neq i} {\rm Im}[(M_D M_D^{\dag})_{ij}^2] 
\left[f\left(\frac{M_j^2}{M_i^2} \right) 
+ g_j\left(\frac{M_j^2}{M_i^2} \right) \right], 
\ee
where $f(x)$ and $g_j(x)$ denote the one loop contribution from 
the vertex and the self-energy, respectively \cite{self}. 
\bea
f(x) &=& \sqrt{x}\ \ln \left(\frac{1+x}{x} \right ), \;\;
g_j(x) \ =\
\frac{(x-1)\sqrt{x}}{(x-1)^2 + 
\left[(M_D M_D^{\dag})_{jj}/(8 \pi v^2 \sin^2 \beta) \right]^2 x} .
\label{self}
\eea
Once we are able to know the values of the lepton asymmetry $\epsilon_i$, 
we have to solve the Boltzmann equations in order to get the 
actual baryon asymmetry, $\eta = n_B/n_\gamma$.  
But we can use an approximate formula \cite{takanishi},
\be
\eta \simeq 3 \times 10^{-3} 
\left(\frac{10^{-3}\; \mathrm{eV}}{\tilde{m}_1} \right) 
\left[
\log \left(\frac{\tilde{m}_1}{10^{-3}\;\mathrm{eV}}\right)
\right]^{-0.6} 
\epsilon_1,
\ee
where 
\be
\tilde{m}_1 = \frac{(M_D M_D^\dag)_{11}}{M_1}. 
\ee
This formula is within a sufficiently good approximation for 
$10^{-2} \;\mathrm{[eV]} < \tilde{m}_1 < 10^{3}\;\mathrm{[eV]}$.

\begin{figure}[p]
\begin{center}
\includegraphics[width=.8\linewidth]{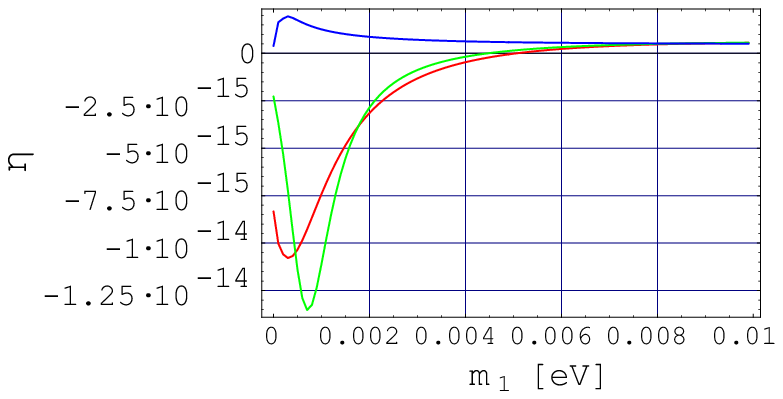}
\end{center}
\caption{
The baryon asymmetry of the universe $\eta$ as a function of the 
lightest neutrino mass $m_1$ [eV] for Model 1. 
Three lines (red, blue, green) correspond to 
the values $\delta = \pi/4,~ \pi/2,~\pi$, respectively. 
In this figure we have fixed $|U_{e3}| = 0.15$ and 
the Majorana phases $\beta$ and $\gamma$ as 
$\beta \ =\ \gamma \ =\ \pi/4$. }
\label{fig1}
\end{figure}

\begin{figure}
\begin{center}
\includegraphics[width=.8\linewidth]{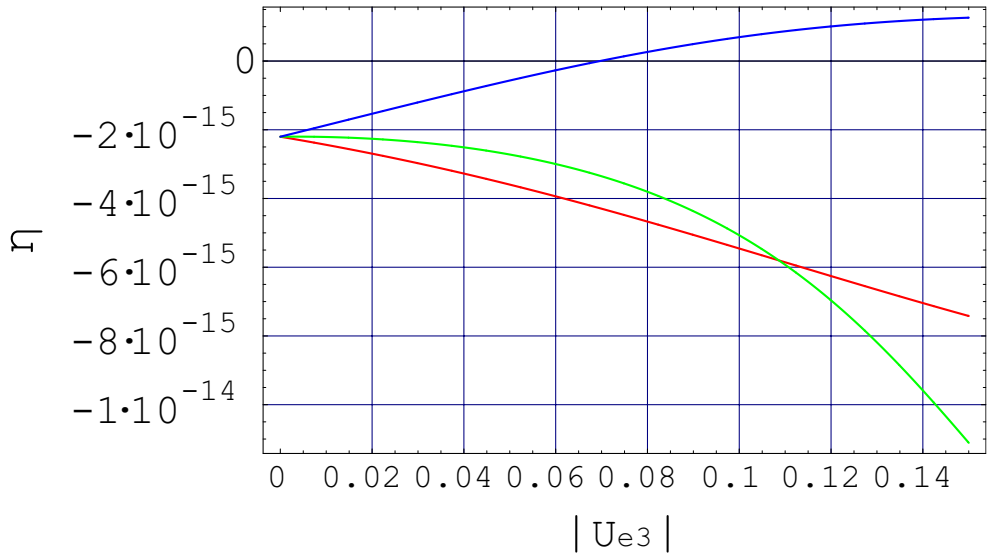}
\end{center}
\caption{
The baryon asymmetry of the universe $\eta$ as a function of 
$|U_{e3}|$ for Model 1. Three lines (red, blue, green) correspond to 
the values $\delta = \pi/4,~ \pi/2,~\pi$, respectively. 
In this figure, we have fixed $m_1 = 10^{-3}$ [eV] 
and the Majorana phases $\beta$ and $\gamma$ as 
$\beta \ =\ \gamma \ =\ \pi/4$. }
\label{fig2}
\end{figure}

\begin{figure}
\begin{center}
\includegraphics[width=.8\linewidth]{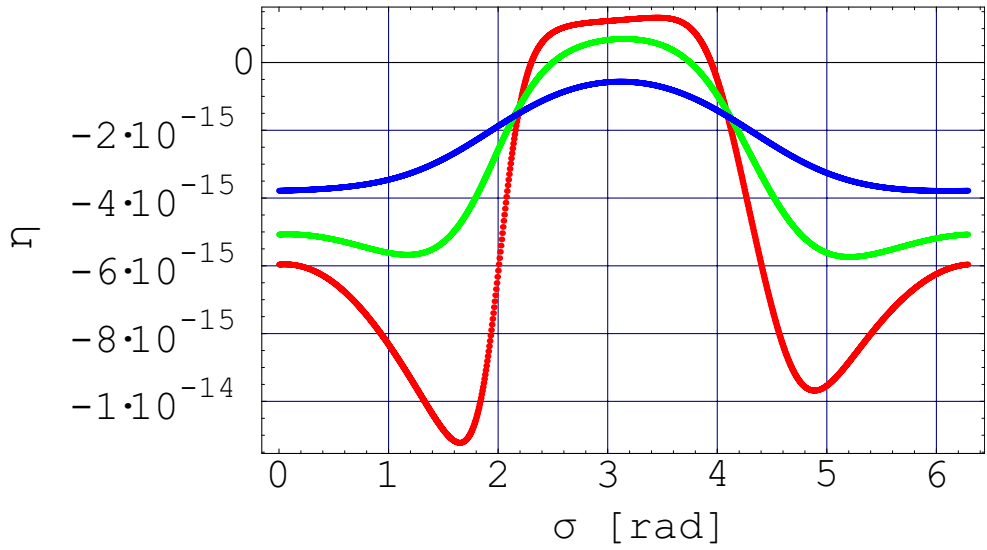}
\end{center}
\caption{
The baryon asymmetry of the universe $\eta$ as a function of 
$\delta$ [rad] for Model 1. Three lines (red, blue, green) correspond to 
the values $|U_{e3}| = 0.15,~ 0.10,~0.05$, respectively. 
In this figure, we have fixed $m_1 = 10^{-3}$ [eV] 
and the Majorana phases $\beta$ and $\gamma$ as 
$\beta \ =\ \gamma \ =\ \pi/4$. }
\label{fig3}
\end{figure}

Now we we show the results of the baryon asymmetry of the universe 
in case of NH, in Fig. \ref{fig1}, \ref{fig2}, and \ref{fig3}. 
The vertical axis represents the predicted baryon asymmetry 
of the universe $\eta$ as a function of the lightest 
neutrino mass $m _1$ [eV] in Fig. \ref{fig1}, 
as a function of $|U_{e3}|$ in Fig. \ref{fig2} and 
as a function of $\delta$ [rad] in Fig. \ref{fig3}. 
In Fig. \ref{fig1}, we have fixed $|U_{e3}| = 0.15$ and 
the Majorana phases $\beta$ and $\gamma$ as 
$\beta \ =\ \gamma \ =\ \pi/4$. 
In Fig. \ref{fig2}, $m_1$ is fixed as $m_1 = 10^{-3}$ [eV] and 
the Majorana phases are the same as Fig. \ref{fig1}. 
For Fig. \ref{fig1} and \ref{fig2}, three lines (red, blue, green) represent 
the values $\delta = \pi/4,~ \pi/2,~\pi$, respectively. 
In Fig. \ref{fig3}, we have fixed $m_1 = 10^{-3}$ [eV] 
and the Majorana phases $\beta$ and $\gamma$ as 
$\beta \ =\ \gamma \ =\ \pi/4$ but varied $|U_{e3}| = 0.15,~ 0.10,~ 0.05$, 
which correspond to the three lines (red, blue, green). 
These results show that Model 1 can not reproduce the observed 
baryon asymmetry in the standard leptogenesis scenario, though 
all possible values of free parameters have not been exhausted. 
But in strictly speaking, any (local) supersymmmetric models have a serious 
problem, so called, gravitino problem \cite{weinberg, Khlopov, kawasaki}. 
That says if we take a na\"ively expected value 
for the garavitino mass $m_{3/2}$ of order the weak sacle 
$m_{3/2} \sim 100$ [GeV] the lifetime is shorter than 1 [sec], as a result 
the primordial gravitino dacays after the big-bang nucleosynthesis (BBN). 
Recently, the updated analysis for the hadronic decay of the gravitino shows 
that the reheating temperature is very strictly constrained as $T_R < 10^{6-8}$ [GeV] 
\cite{kawasaki}.
Therefore the results of our Model 1 for the baryon asymmetry of the universe should not taken so seriously. 

\begin{figure}[p]
\begin{center}
\includegraphics[width=.8\linewidth]{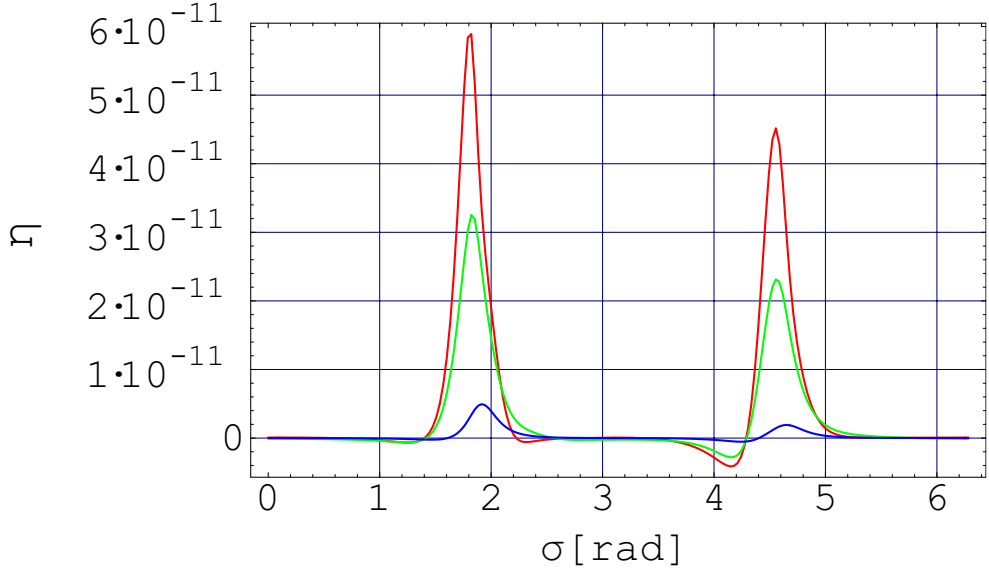}
\end{center}
\caption{
The baryon asymmetry of the universe $\eta$ as a function of 
$\sigma$ [rad] in Model 2. 
Three lines (red, blue, green) correspond to 
the values $\delta = \pi/4,~ \pi/2,~\pi$, respectively. 
In this figure, we have fixed 
$m_1 = 10^{-3}$ [eV] and $|U_{e3}|=0.15$, 
and the Majorana phases as $\beta \ =\ \gamma \ =\ \pi/4$. }
\label{fig4}
\end{figure}

\begin{figure}[p]
\begin{center}
\includegraphics[width=.8\linewidth]{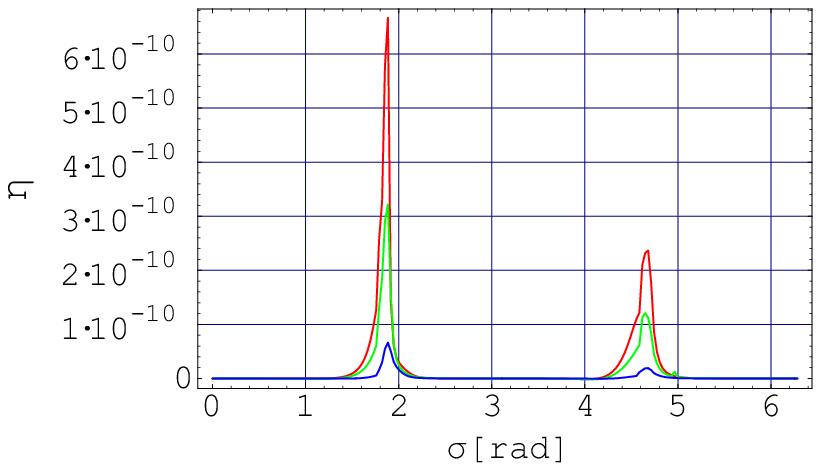}
\end{center}
\caption{
Same as Fig.~\ref{fig4}, but for the Majorana phases 
$\beta \ =\ \gamma \ =\ \pi/2$. }
\label{fig5}
\end{figure}

\begin{figure}
\begin{center}
\includegraphics[width=.8\linewidth]{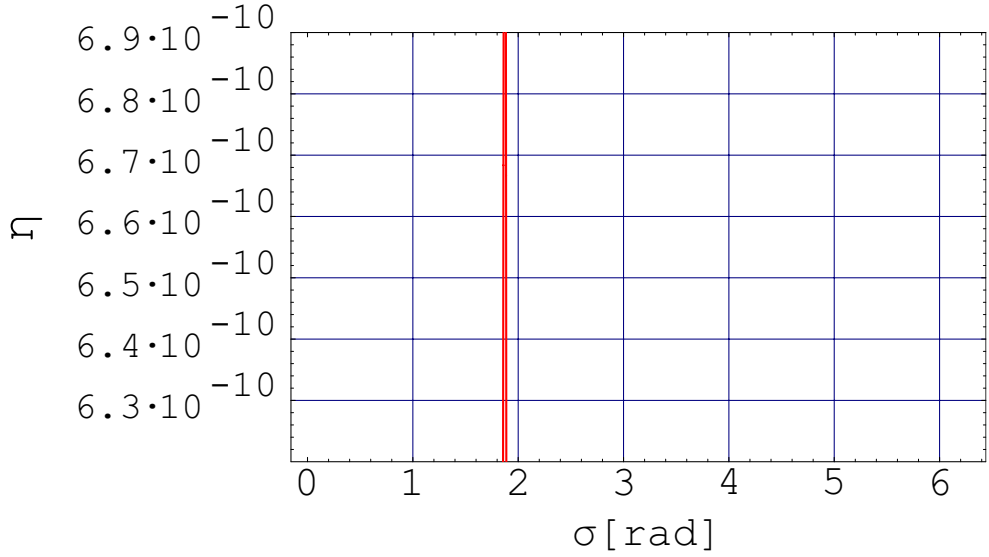}
\end{center}
\caption{The experimentally allowed region in Fig.~\ref{fig5} 
are zoomed in for $\delta \ =\ \pi/4$. }
\label{fig6}
\end{figure}

\begin{figure}[p]
\begin{center}
\includegraphics[width=.8\linewidth]{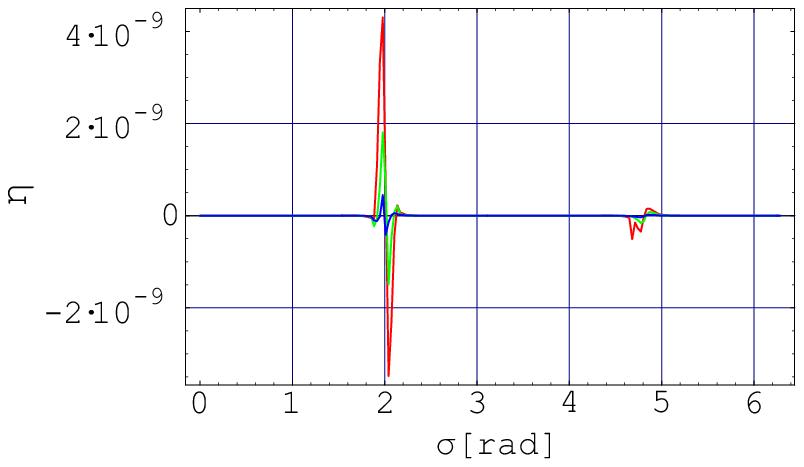}
\end{center}
\caption{
Same as Fig.~\ref{fig4}, but for the Majorana phases 
$\beta \ =\ \gamma \ =\ \pi$. }
\label{fig7}
\end{figure}

On the other hand, in Fig. \ref{fig4}, \ref{fig5}, \ref{fig6} and \ref{fig7} 
we show the results of the baryon asymmetry of the universe 
in case of Model 2. 
The vertical axis represents the predicted baryon asymmetry 
of the universe $\eta$ as a function of the CP-phase 
$\sigma$ [rad]. 
In Fig. \ref{fig4}, we have fixed $m_1 = 10^{-3}$ [eV] and 
$|U_{e3}| = 0.15$, and Majorana phases $\beta$ and $\gamma$ as 
$\beta \ =\ \gamma \ =\ \pi/4$. 
In Fig. \ref{fig5}, we have adopted the same parameters as Fig. \ref{fig4} 
except for the Majorana phases $\beta \ =\ \gamma \ =\ \pi/2$. 
Fig. \ref{fig6} is the zoomed up picture of the experimentally 
allowed parameters region of Fig. \ref{fig5} for the Dirac CP-phase 
$\delta = \pi/4$. 
Fig. \ref{fig7} is the same diagram as Fig. \ref{fig4} except for the 
Majorana phases $\beta \ =\ \gamma \ =\ \pi$. 
You can see that this model can reproduce the observed 
baryon asymmetry within the standard leptogenesis scenario. 

\begin{figure}[p]
\begin{center}
\includegraphics[width=.8\linewidth]{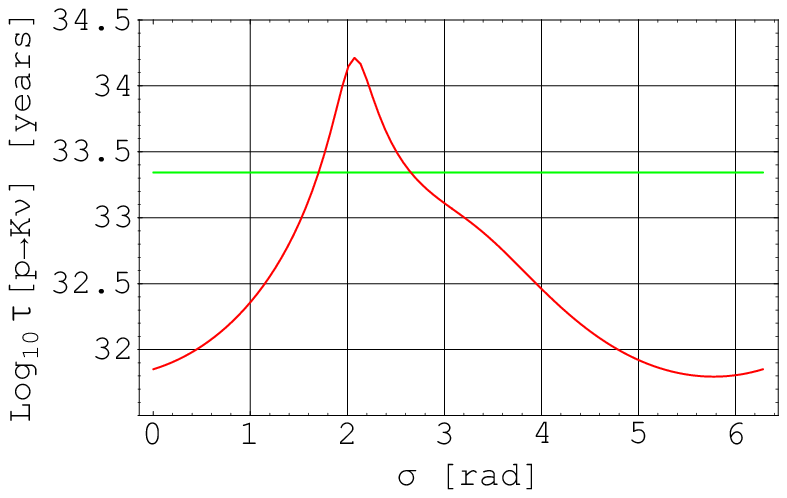}
\end{center}
\caption{
The proton lifetime $\tau(p \to K^+ \bar{\nu})$ in a mode 
$p \to K^+ \bar{\nu}$ as a function of $\sigma$ [rad] for Model 2. 
Though in general we can vary the parameters in the Higgs potential 
as possible, we pick up one solution satisfying the experimental constraint. 
The horizontal line indicates the current experimental lower bound. }
\label{fig8}
\end{figure}

Since Model 2 has passed the test for the leptogenesis, it is possible 
to go further steps, e.g. analysis of the proton lifetime 
$\tau(p \to K^+ \bar{\nu})$ in a mode $p \to K^+ \bar{\nu}$.  
The predicted values as a function of $\sigma$ [rad] 
are depicted in Fig. \ref{fig8}. 
Though in general we can vary the parameters in the Higgs potential 
as possible, we pick up one solution satisfying the experimental constraint. 
In concrete, we used the $2 \times 2$ color-triplet Higgsino 
mass matrix \cite{Detailed} of the following form, 
\be
M_C  = M_G {\bf I}_2 \times 
\left(
\begin{array}{cc}
\begin{array}{c}
1 +i  \\
- 1 + i
\end{array}
\begin{array}{c}
1 +i \\
-1 - i
\end{array} 
\end{array}
\right), 
\ee
which corresponds to the parameters with 
$\theta \ =\ \varphi \ =\ \varphi^\prime \ =\ \pi/4$ 
in the notation of Ref. \cite{Detailed}. 
The horizontal line indicates the current experimental lower bound. 
Therefore, these results show that there exists a parameter region 
which simultaneously reproduces a required baryon asymmetry of the universe 
and satisfies the current experimental bound on the proton lifetime: 
\bea
\eta|_{\rm exp.} = (6.2-6.9) \times 10^{-10}, 
\\
\tau(p \to K^+ \bar{\nu})|_{\rm exp.}  > 2.2 \times 10^{33} ~{\rm [years]}.
\eea
\section{Conclusion}
We have proposed a phenomenologically viable ${\rm SO}(10)$ grand unification 
model which use a set of Higgs superfields 
${\bf 10 + \overline{16} + 16 + 45}$. 
In this framework, the data fitting of the 
charged fermion mass matrices have been re-examined. This model indeed can 
fit the low-energy experimental data relating the charged 
fermion masses and mixings. As for the neutrino sector, we have inputted 
the neutrino oscillation data to constrain right-handed Majorana neutrino 
mass matrix. Then the unknown parameters are used to fit the leptogenesis 
and for the proton decay. In such a framework, two models, 
Model 1 and Model 2, have been considered. The former is effectively equal 
to the model with the Yukawa couplings to the ${\bf 10 + 120}$ Higgses 
and the latter to the model with the Yukawa couplings 
to the ${\bf 10 + \overline{126}}$ Higgses with respect to charged Fermions. 
It should be remarked in Model 2 that $M_R$ has the different origin from 
that of the charged fermion mass matrices unlike the conventional 
minimal ${\rm SO}(10)$ model \cite{matsudaetal} \cite{F-O}. 
Then it has been found that our model (Model 2) is consistent with 
all experimental constraints, especially, the required baryon asymmetry 
of the universe $\eta$ can be produced, keeping in mind the proton 
lifetime constraint. Though Model 1 does not produce the sufficient 
baryon asymmetry of the universe within the standard framework of the 
leptogenesis, there is a serious problem in any SUSY models, so called, 
gravitino problem \cite{weinberg, Khlopov, kawasaki}. The decay products of the primodial 
gravitino destroy the produced baryon asymmetry of the universe 
and need further study to consider some extensions 
of the standad leptogensis scenario, 
such as Affleck-Dine scenario \cite{AD}.

\section*{Acknowledgments}
T.K. would like to thank K.S. Babu
 for his hospitality at Oklahoma State University.
The work of T.K. is supported by the Research
Fellowship of the Japan Society for the Promotion of Science (\#1911329).



\begin{thebibliography}{99}
\bibitem{unification}
U. Amaldi, W.de Boer, 
and H. F{\"u}rstenau, Phys. Lett. {\bf B260},447, (1991); \\
P. Langacker and M. Luo, Phys. Rev. {\bf D44}, 817 (1991).  
%
\bibitem{seesaw}
T. Yanagida, in Proceedings of the workshop 
on the Unified Theory and Baryon Number in the Universe, 
edited by O. Sawada and A. Sugamoto (KEK, Tsukuba, 1979); 
%
P.~Ramond, CALT-68-709, Feb 1979. 21pp.
Invited talk given at Sanibel Symposium, Palm Coast, Fla., Feb 25 -
Mar 2, 1979. Published in *Paris 2004, Seesaw 25* 265-280,
e-Print: hep-ph/9809459;
%
R.N. Mohapatra and G. Senjanovi\'c, 
Phys.\ Rev.\ Lett.\ {\bf 44}, 912 (1980). 
%
\bibitem{string}
K.R. Dienes, J. March-Russell, 
Nucl. Phys. {\bf B479}, 113, (1996); 
K.R. Dienes, Nucl. Phys. {\bf B488}, 141, (1997); 
see also a review article, K.R. Dienes, 
Phys. Rept. {\bf 287}, 447 (1997). 
%
\bibitem{alternative}
H.~S.~Goh, R.~N.~Mohapatra and S.~Nasri,
  Phys.\ Rev.\  D {\bf 70}, 075022 (2004)
  [arXiv:hep-ph/0408139];
B.~Dutta, Y.~Mimura and R.~N.~Mohapatra,
  Phys.\ Rev.\ Lett.\  {\bf 100}, 181801 (2008)
  [arXiv:0712.1206 [hep-ph]].
%
\bibitem{babu}
K.S. Babu and R.N. Mohapatra, 
Phys. Rev. Lett. {\bf 70}, 2845 (1993). 
%
\bibitem{matsudaetal}
K.~Matsuda, Y.~Koide and T.~Fukuyama, 
Phys.\ Rev.\ D {\bf 64}, 053015 (2001) 
[arXiv:hep-ph/0010026]; 
K.Matsuda, Y.Koide, T.Fukuyama and H.Nishiura, 
Phys. Rev. {\bf D65}, 033008 (2002) [arXiv:hep-ph/0108202]. 
\bibitem{F-O}
T.~Fukuyama and N.~Okada, 
JHEP {\bf 0211}, 011 (2002) [arXiv:hep-ph/0205066].
%
\bibitem{mohapatra}
B. Bajc, G. Senjanovi\'c and F. Vissani, 
Phys. Rev. Lett. {\bf 90}, 051802 (2003); 
H. S. Goh, R.N. Mohapatra, S. Nasri, Siew-Phang Ng, 
Phys. Lett. {\bf B587} 105 (2004); 
B. Dutta, Y. Mimura, and R.N. Mohapatra, 
[arXiv:hep-ph/0406262].
%
\bibitem{non-ren}
L.J. Hall, R. Rattazzi and U. Sarid, Phys. Rev. {\bf D50}, 7048 (1994); 
R. Rattazzi and U. Sarid, Phys. Rev. {\bf D53}, 1553 (1996); 
%
G. Anderson, S. Dimopoulos, L.J. Hall, S. Raby and G. D. Starkman, 
Phys. Rev. {\bf D49}, 3660 (1994); 
%
L.J. Hall and S. Raby, Phys. Rev. {\bf D51}, 6524, (1995); 
%
%
C.H. Albright, K.S. Babu and S.M. Barr, Phys. Rev. Lett. {\bf 81}, 1167 
(1998); 
C.H. Albright and S.M. Barr, Phys. Lett. {\bf B461}, 218 (1999), 
Phys. Rev. Lett. {\bf 85}, 244 (2000); Phys. Rev. {\bf D62}, 093008 (2000); 
%
Q. Shafi and Z. Tavartkiladze, Phys. Lett. {\bf B482}, 145 (2000); 
%
K.S. Babu, J.C. Pati and F. Wilczek, Nucl. Phys. {\bf B566}, 33 (2000). 
%
\bibitem{FIKMO2}
T.~Fukuyama, A.~Ilakovac, T.~Kikuchi, S.~Meljanac and N.~Okada,
  Eur.\ Phys.\ J.\  C {\bf 42}, 191 (2005)
  [arXiv:hep-ph/0401213];
  T.~Fukuyama, A.~Ilakovac, T.~Kikuchi, S.~Meljanac and N.~Okada,
  J.\ Math.\ Phys.\  {\bf 46}, 033505 (2005)
  [arXiv:hep-ph/0405300];
  T.~Fukuyama, A.~Ilakovac, T.~Kikuchi, S.~Meljanac and N.~Okada,
  Phys.\ Rev.\  D {\bf 72}, 051701(R) (2005)
  [arXiv:hep-ph/0412348]. 

\bibitem{Chang:2004pb}
  D.~Chang, T.~Fukuyama, Y.~Y.~Keum, T.~Kikuchi and N.~Okada,
  Phys.\ Rev.\  D {\bf 71}, 095002 (2005)
  [arXiv:hep-ph/0412011].

\bibitem{PDG}
K. Hagiwara {\it et al.} [Particle Data Group Collaboration], 
Phys. Rev. {\bf D66}, 010001 (2002).  
\bibitem{Detailed}
T. Fukuyama, A. Ilakovac, T. Kikuchi, S. Meljanac and N. Okada, 
JHEP {\bf 0409}, 052 (2004). 

\bibitem{petcov}
A. Bandyopadhyay, S. Choubey, S. Goswami, S.T. Petcov, D.P. Roy, 
[arXiv:hep-ph/0406328]. 

\bibitem{FY}
M. Fukugita, T. Yanagida, 
Phys. Lett. {\bf B 174}, 45 (1986). 

\bibitem{self}
M. Flanz, E.A. Paschos, U. Sarkar, 
Phys. Lett. {\bf B 345}, 248 (1995); 
Erratum-ibid. {\bf B382}, 447 (1996); 
L. Covi, E. Roulet, F. Vissani, 
Phys. Lett. {\bf B 384}, 169 (1996); 
W. Buchmuller, M. Plumacher, 
Phys. Lett. {\bf B 431}, 354 (1998). 

\bibitem{takanishi}
E.W.~Kolb and M.S.~Turner, 
{\it The Early Universe}, Addison-Wesley (1990). 

\bibitem{weinberg}
S.~Weinberg, Phys.\ Rev.\ Lett.\ {\bf 48}, 1303 (1982).

\bibitem{Khlopov}
M. Yu. Khlopov and A. D. Linde, Phys.\ Lett. {\bf B~138} (1984) 265; 
I.~V.~Falomkin, G.~B.~Pontecorvo, M.~G.~Sapozhnikov, M.~Yu.~Khlopov, 
F.~Balestra and G.~Piragino, 
Nuovo Cim. {\bf A~79} (1984) 193 [Yad.\ Fiz.\  {\bf 39} (1984) 990]; 
M.~Yu.~Khlopov, Yu.~L.~Levitan, E.~V.~Sedelnikov and I.~M.~Sobol, 
Phys.\ Atom.\ Nucl.\  {\bf 57} (1994) 1393 [Yad.\ Fiz.\  {\bf 57} (1994) 1466].

\bibitem{kawasaki}
M.~Kawasaki and T.~Moroi,
  Prog.\ Theor.\ Phys.\  {\bf 93}, 879 (1995)
  [arXiv:hep-ph/9403364];
M.~Kawasaki, K.~Kohri and T.~Moroi,
  Phys.\ Lett.\  B {\bf 625}, 7 (2005)
  [arXiv:astro-ph/0402490];
M.~Kawasaki, K.~Kohri and T.~Moroi,
  Phys.\ Rev.\  D {\bf 71}, 083502 (2005)
  [arXiv:astro-ph/0408426].

\bibitem{AD}
I. Affleck, M. Dine, Nucl. Phys. {\bf B249}, 361 (1985). 
\end{thebibliography}
\end{document}